\documentclass[preprint,12pt,authoryear]{elsarticle}

\usepackage{amssymb}
\usepackage{amsmath}
\usepackage{float}
\usepackage{booktabs}
\usepackage{url}

\journal{Chaos, Solitons \& Fractals}

\begin{document}

\begin{frontmatter}



\title{Nonlinear dynamics of information overload: Impact on source localization in complex networks} 


\author[label1]{Ignacy Czajkowski} 
\author[label1]{Robert Paluch}

\affiliation[label1]{organization={Faculty of Physics, Warsaw University of Technology},
            addressline={Koszykowa 75}, 
            city={Warsaw},
            postcode={00662}, 
            country={Poland}}

\begin{abstract}
Source localization in complex networks is a rapidly advancing field with numerous real-world applications, including determining the source of misinformation.
In this work, we model information spread across several real-world and synthetic complex networks using our Generalized Fractional Susceptible-Infected-Recovered (GFSIR) model, which incorporates the information overload (IOL) phenomenon.
Then, we use Pearson's correlation algorithm to identify information sources in these networks and investigate how information overload affects localization quality.
Numerical simulations have shown that localization effectiveness decreases with the parameter $\alpha$, which controls the strength of the IOL, and increases with the spreading rate $\beta$.
Our comparison across various topologies reveals that localization is generally more effective in synthetic structures, with Erd\H{o}s-R\'{e}nyi networks exhibiting greater resilience to IOL than Barab\'{a}si-Albert models.
Furthermore, we identified a critical reversal in the impact of network density: while a higher average degree enhances localization when IOL is negligible, less dense networks perform better under strong overload.
This phenomenon represents a significant departure from the behavior observed in standard epidemic models.
\end{abstract}

\begin{highlights}
 \item Information overload (IOL) significantly hinders information source localization.
 \item High spreading rates effectively weaken the negative impact of the IOL effect.
 \item Network density and clustering coefficients enhance the strength of IOL.
\end{highlights}

\begin{keyword}
Complex Networks \sep Generalized Fractional Susceptible-Infected-Recovered model \sep Information Overload \sep Source Localization 


\end{keyword}

\end{frontmatter}









\section{Introduction}
In modern society, we increasingly rely on complex systems that permeate various aspects of our daily lives \citep{Artime2024}.
Prominent examples include social networks \citep{networks_in_social}, which facilitate communication and information sharing; financial markets \citep{networks_in_finance}, where interconnected transactions influence global economies; and transportation networks, which enable the efficient movement of goods and people through complex logistics and infrastructure \citep{Ding2019, Lordan2020}.
Given their ubiquity and significant impact, there is a growing need to understand the underlying dynamics of these systems.
To address this, researchers have developed models based on complex networks, where nodes represent individual agents and edges denote the connections between them \citep{complex_networks_models}.

The network-based approach has proven highly effective, with numerous studies demonstrating that interactions among nodes can profoundly influence the behavior and functionality of the entire system.
Notable examples include the dissemination of risk in financial networks \citep{risk_dis_in_finance}, the slowdown in social and economic development during disease outbreaks \citep{delayed_development}, and the spread of rumors and misinformation \citep{rumors_and_misinf_spread, Ji2023, Gosh2024}.

Recently, source localization has emerged as a critical challenge in today's digital landscape, where access to vast amounts of information is unprecedented.
However, this increased accessibility also presents the difficulty of assessing the information's credibility.
Accurately identifying the source of information can play a pivotal role in evaluating its reliability, thereby helping to curb the spread of misinformation.
Source localization is equally important in public health, where early detection of epidemic outbreaks can facilitate more efficient preventive measures, potentially mitigating the scope and severity of disease spread \citep{epidemic_spreading}.

Traditional approaches to source localization have primarily focused on developing algorithms that exploit network topology and observed infection patterns \citep{zhu2016information, Lu2025, Huang2025, Fraszczak2025}.
However, these methods typically assume simplified propagation models that do not account for the nonlinear dynamics of real-world information diffusion, particularly the phenomenon of information overload (IOL).
In reality, individuals and nodes in networks have limited cognitive capacity and attention, leading to nonlinear responses in which the probability of accepting new information decreases with exposure \citep{eppler2004conceptual, bawden2008information}.
This nonlinear relationship between information exposure and acceptance represents a fundamental departure from classical epidemic models and introduces complex system dynamics that can dramatically alter propagation patterns.

The recent years 2025-2026 have witnessed a surge in the application of Large Language Models (LLMs) for combating misinformation through sophisticated semantic filtering and content-based verification \citep{Huang2025b}. 
However, while these AI-driven approaches are highly effective at the semantic layer, they often overlook the topological layer of information dissemination.
Even with perfect content identification, the network's physical constraints -- most notably information overload -- distort propagation patterns, rendering source localization a purely structural challenge that language-centric models cannot resolve.

Recent studies have begun to explore the implications of information overload in various contexts \citep{inf_ovrload_hist}, but its impact on source localization algorithms remains largely unexplored.
Understanding how nonlinear dynamics and phase-transition-like behaviors emerge from information overload (IOL) is critical for developing robust localization methods that can operate effectively in real-world scenarios characterized by complex system properties.
This paper addresses this gap by introducing the Generalized Fractional SIR (GFSIR) model, a novel extension of the classical SIR epidemic model that incorporates nonlinear information overload dynamics.

The key contributions of our study are:
\begin{itemize}
    \item Introducing a generalized version of the FSIR model (GFSIR) to capture the nonlinear dynamics of information spread under information overload (IOL) conditions.
    \item Analyzing the impact of IOL on the source localization effectiveness for various spreading and detection parameters and for different synthetic and real-world networks.
    \item By concluding simulations, we found out that there exists a visible impact of IOL on source localization.
    \item We presented an explanation of the impact of IOL on source localization and differences in impact across networks with varying average degree.
\end{itemize}

The rest of the paper is organized as follows.
Section 2 discusses previous research on source localization in complex networks and on the information overload phenomenon.
Section 3 describes the GFSIR information-propagation model, the correlation-based source localization algorithm, and evaluation metrics for localization effectiveness.
Section 4 presents the conditions of the conducted simulations and provides an overview of the simulation results.
Section 5 includes conclusions and future research directions.

\section{Related works}

The development of source localization algorithms is a rapidly advancing field, characterized by a wide range of methods and data acquisition requirements for the spreading process \citep{source_det_alg_clas, Ali2025, Hu2025}.
Most established localization algorithms are observer-based, with specific nodes acquiring data during the spreading process for later use in detection, as demonstrated in the earliest well-known algorithms introduced by \cite{ptv_algorithm} and \cite{s_z_algorithm}.
One of the most robust localization algorithms to this day is Pearson's correlation algorithm \citep{corelation_algorythm}.
Its strength stems from its efficacy, even with a limited number of observers and relatively low computational complexity \citep{loc_alg_comp}.
Furthermore, Pearson's correlation algorithm does not require any additional details about the information-spreading process, such as the identification of nodes from which observers obtained the information, thereby making it more applicable to real-world problems.

Researching the effectiveness of source localization typically involves using computational simulations to model information propagation in networks \citep{source_det_in_soc_net}.
These simulations rely on a network model, an information-propagation model, and an algorithm to evaluate the source after the information-spreading process.
Commonly used spread models include well-known epidemiological models, such as the SI \citep{si_model}, SIS \citep{sis_model}, or SIR \citep{sir_model_in_inf} models.
Research in this field encompasses both single-source and multi-source spread \citep{multi_source_det, multi_source_det2}, with a focus on various network types, including synthetic and real networks.
The effectiveness of localization is assessed by comparing the true information source with the node (or nodes) identified by the localization algorithm, using metrics such as precision or ranking \citep{source_det_in_soc_net}.

The concept of information overload (IOL), also referred to as infobesity \citep{information_overload_explain1}, infoxication, or information anxiety \citep{information_overload_explain2}, refers to the challenge of processing and comprehending excessive amounts of information.
Coined in the book \textit{"The Managing of Organizations: The Administrative Struggle"} \citep{inf_ovr_book}, this phenomenon occurs when the input of information surpasses a system's capacity to process it \citep{inf_ovr_in_age}.
Researchers across various disciplines investigate information overload in terms of classification, its impact on human health and behavior, frequency of occurrence, and methods to mitigate its adverse effects \citep{Holyst2024}.

An example of research on the impact of information overload on behavior is a study by \cite{sso_framework}, which investigates the effects of IOL on human exhaustion and discontinuation of social media use.
They introduced the Stressor-strain-outcome (SSO) framework and found a correlation between information overload and social media fatigue.

\cite{soc_econ_stat_inf_ovr} investigated the impact of an infodemic on vaccine hesitancy, considering socioeconomic and cultural factors.
The research data were based on information available online about the COVID-19 epidemic, and the infodemic was defined as a combination of misinformation and information overload.
Based on the distinction between the proposed infodemic factors, they found that people of lower economic status and younger individuals are more susceptible to overload. 

Another type of study focuses on ways to surpass information overload to mitigate its adverse effects.
To overcome overload, there was a need for accurate classification of information by relevance.
\cite{ml_for_inf_relevance_class} proposed machine-learning techniques for information relevance classification.
They used a convolutional neural network and a random forest classifier, training the models on data from social media posts about disasters and emergencies.    

To capture the effects of user information overload on propagation dynamics, specialized models have been developed.
A prominent example is the Fractional Susceptible-Infected-Recovered (FSIR) model, proposed by \cite{fsir_model} for analyzing the dissemination of news on social media.
Validation against real-world data confirms that, unlike classical epidemic models, the FSIR framework correctly predicts a finite threshold for information to become viral.
This model is based on the central premise that an individual's capacity to process information is limited by their attentional resources.

Recent state-of-the-art research (e.g., \cite{He2026}) explores hybrid frameworks that combine machine learning with graph theory to track digital rumors.
Nevertheless, a significant gap remains regarding the impact of user \textit{infobesity} on the reliability of these methods.
While standard epidemic models such as SI or SIR have been widely adopted \citep{Ji2023}, their classical versions fail to capture the fractional-order slowing of information flow observed in high-density environments.
Our approach builds on this limitation by focusing on the interplay between network architecture and the fractional dynamics of overloaded agents, a direction increasingly recognized in 2026 as vital for robust source detection.

\section{Preliminaries}

\subsection{Information propagation model}
The GFSIR (Generalized Fractional Susceptible-Infected-Recovered) model is an extension of the SIR epidemic framework, where nodes transition between three states: Susceptible (\(S\), node has not received information), Infected (\(I\), node has received and is spreading information), and Recovered (\(R\), node has lost interest and stopped spreading).
The dynamics are governed by a spreading rate \(\beta  \in [0,1]\) and a recovery rate \(\gamma \in [0,1]\).
As this study focuses on localizing the source of viral information (e.g., rumors, malware) during its initial, rapid spread, we assume users do not lose interest in it.
We therefore simplify the dynamics by setting the recovery rate \(\gamma=0\).

In the standard SIR model, the \(S \to I\) transition probability depends on the number of infected neighbors, \(k_i(t)\).
Each of these \(k_i(t)\) neighbors transmits the information with an independent probability \(\beta\).
The cumulative probability that node \(i\) becomes infected at time \(t+1\) is the complement of it remaining uninfected by all \(k_i(t)\) neighbors:
\begin{equation}
            p_{i,S\rightarrow I}(t) = 1 - {(1-\beta)}^{k_i(t)}
\label{si_state_eq}            
\end{equation}

A key limitation of the SIR model is its failure to account for information overload.
The FSIR and GFSIR models address this by positing that users have limited attention.
They modify the transmission mechanism: the individual probability of infection is no longer a constant \(\beta\) but is \emph{fractionalized} by the number of simultaneous exposures.
In the GFSIR model, this probability becomes \(\beta/k_i(t)^\alpha\), where \(\alpha\geq 0\) is a parameter quantifying the strength of the information overload.
This generalization distinguishes GFSIR from the original FSIR model, which assumes \(\alpha=1\).
Substituting this fractionalized probability into the cumulative infection formula yields:
\begin{equation}
            p_{i,S\rightarrow I}(t) = 1 - {(1-\frac{\beta}{{k_i(t)}^{\alpha}})}^{k_i(t)}
\label{fsir_state_eq}            
\end{equation}.

The GFSIR model introduces fundamental nonlinear dynamics into information propagation through its IOL mechanism, representing a significant departure from classical linear epidemic models.
This nonlinearity manifests in the relationship between information exposure and acceptance probability, capturing the complex cognitive processes underlying information overload in real-world networks.
To understand the nonlinear behavior, consider the infection probability in Equation~\ref{fsir_state_eq}.
When \(\alpha > 0\), the model exhibits a nonlinear response to the number of infected neighbors \(k_i(t)\).
Unlike the standard SIR model, where infection probability increases monotonically with \(k_i(t)\), the GFSIR model introduces a saturation effect: as \(k_i(t)\) grows, the per-contact transmission probability \(\beta/k_i(t)^\alpha\) decreases, leading to diminishing returns in overall infection probability.
This nonlinear relationship can be analyzed by examining the limiting behavior of \(p_{i,S\rightarrow I}(t)\) as \(k_i(t) \to \infty\).
For \(\alpha > 1\), the infection probability approaches zero, indicating complete saturation where additional exposures provide negligible benefit.
For \(\alpha = 1\), the probability converges to \(1 - e^{-\beta}\), representing a critical threshold behavior.
For \(0 < \alpha < 1\), the probability approaches 1, but at a sublinear rate compared to the standard SIR model.
These distinct regimes demonstrate phase-transition-like behavior in the system's response to information overload.
	
The nonlinearity parameter \(\alpha\) thus controls the strength of cognitive saturation effects.
When \(\alpha = 0\), the model reduces to the standard SIR framework with linear dynamics.
As \(\alpha\) increases, the nonlinear IOL effect becomes more pronounced, fundamentally altering the propagation dynamics and giving rise to emergent behaviors characteristic of complex systems.
This includes the emergence of effective spreading thresholds, topology-dependent phase transitions, and heterogeneous propagation patterns that depend on local network structure.

\subsection{Correlation-based source detection}
In our work, we utilize a correlation-based algorithm for information source localization \citep{corelation_algorythm}.
This method relies on a set of \emph{observers}, specific nodes in the network designated to record their activation times (i.e., the time steps at which they received the information).
The algorithm assigns a score \(s_i\) to each node \(i\) that represents the Pearson correlation between the two vectors.
The first vector, $\vec{t}$, contains the activation times recorded by \(K\) observers:
\begin{equation}
            \vec{t} = [t_{1}, t_{2}, ..., t_{K}]
\label{time_vector_eq}            
\end{equation}
The second vector, $\vec{d_i}$, contains the shortest path distances between the candidate node \(i\) and each observer:
\begin{equation}
            \vec{d_i} = [d(i,o_1), d(i,o_2), ..., d(i,o_K)].
\label{distance_vector_eq}            
\end{equation}
where \(d(i,o_j)\) is the length of the shortest path between node $i$ and observer $o_j$.
The score for node \(i\) is then calculated as:
\begin{equation}
    s_i = \text{cor}(\vec{t}, \vec{d_i})
\label{corelation_eq}    
\end{equation}
The node with the highest score \(s_i\) is identified as the predicted source.
If a tie occurs, all nodes with the maximum score are considered potential sources.

\subsection{Evaluation metrics}
Source localization effectiveness is evaluated using two well-known measures.
The first of such measures is precision $p \in [0,1]$, which is the fraction of the number of correctly indicated sources to the number of all indicated sources:
\begin{equation}
    p = \frac{TP}{TP+FP}
\label{prec_def_eq}    
\end{equation}
where $TP$ stands for true positive and describes correctly indicated sources, and $FP$ stands for false positive and describes the number of nodes incorrectly indicated as sources. 

The second measure is ranking, which is determined by the position of the true source score $s_{true}$ in a list of all nodes' scores arranged in descending order.
In the case of a tie between $s_{true}$ and any other scores, the ranking is assigned to the last position in the tie.
For instance, if $s_{true} = 0.93$, and the top four scores are 0.97, 0.93, 0.93, 0.76, the ranking would be 3.

\section{Results}
The results presented in this section were obtained through a standardized simulation process.
For each run, a network was loaded or generated, a random source node was selected, and a set of observers (density \(\rho\)) was sampled.
The GFSIR model, with specified \(\alpha\) and \(\beta\) parameters, was used to simulate the information spread, during which observer activation times ($\vec{t}$) were recorded.
Subsequently, Pearson's correlation algorithm was employed to identify potential sources.
Finally, the algorithm's output was compared against the true source to evaluate its average precision and ranking.

To investigate the impact of information overload on source localization, simulations were conducted for various combinations of GFSIR model parameters $\alpha \in [0.0;3.0]$ and $\beta \in [0.1;0.9]$.
Information propagation was simulated on three real-world networks and two network models, Erd\H{o}s-R\'{e}nyi (E-R) and Barab\'{a}si-Albert (B-A).
The first real network, Uni-Email, depicts the email communication connections among staff at Rovira i Virgili University \citep{real_networks, email_network1}.
The second, UCIrvine, represents the correspondence of an online student community from the University of California \citep{real_networks, ucirving_network}.
The last network, Infectious, was built from face-to-face interactions during an exhibition in Dublin \citep{real_networks, infectious_network}.
All networks were treated as undirected, unweighted graphs, and nodes with degree $k=0$ were excluded.
Their parameters are detailed in Table \ref{networks_table}.
For all parameter combinations and network types, simulations were run for three different observer densities $\rho \in \{10\%, 15\%, 20\%\}$.
Please refer to \ref{app1} for more details about simulations.

\begin{figure}[H]
            \centering
            \includegraphics[width=0.85\textwidth]{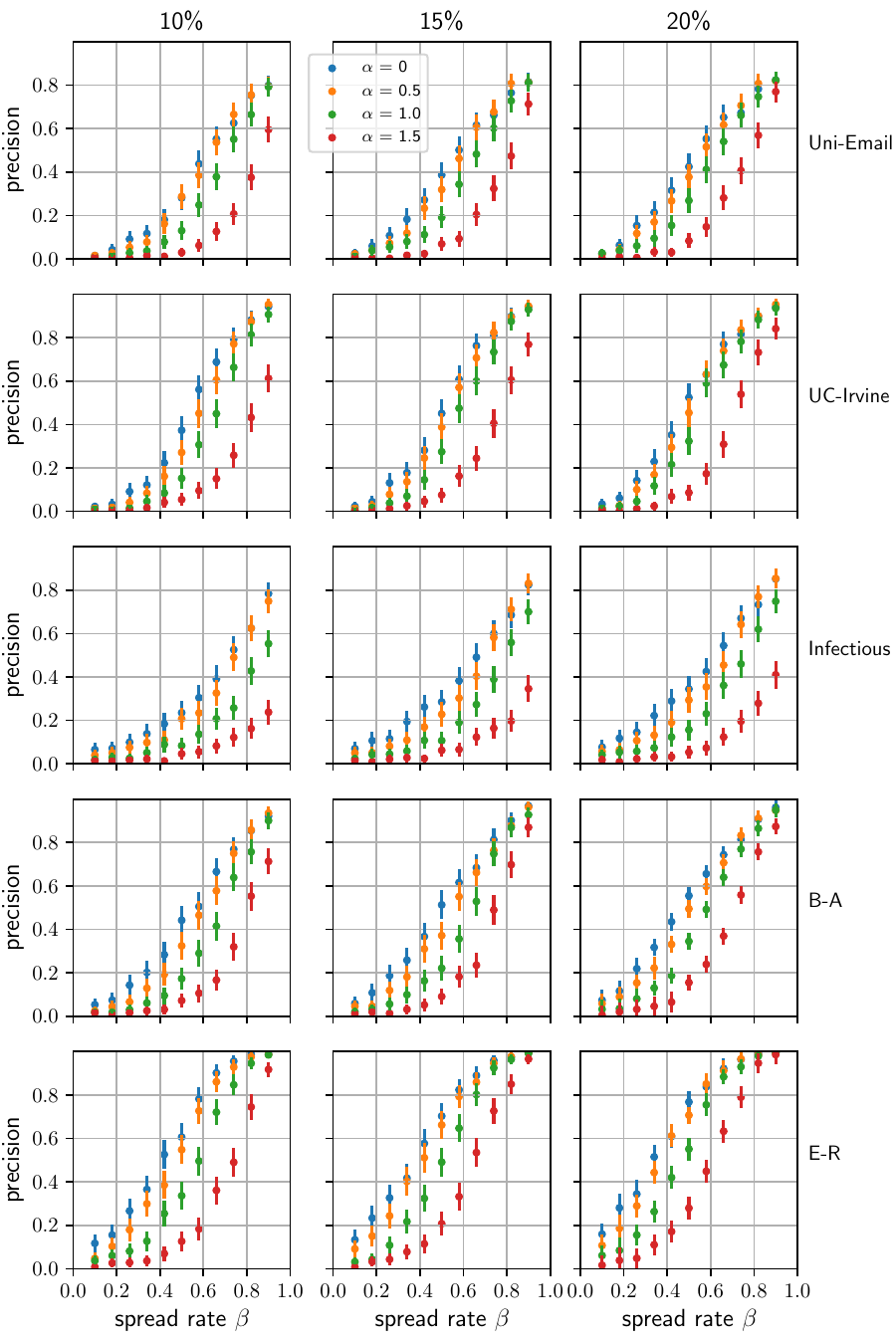}
            \caption{Average precision of source localization increases with the infection rate $\beta$ and decreases with the IOL strength $\alpha$ (indicated in different colors).
            Rows correspond to different networks, and columns correspond to different observer density $\rho$.
            Each point is the average of 500 simulations.
            The error bars represent the extended standard error \((k=3)\).}
            \label{prec_beta_plot}
\end{figure}

\begin{figure}[H]
            \centering
            \includegraphics[width=0.85\textwidth]{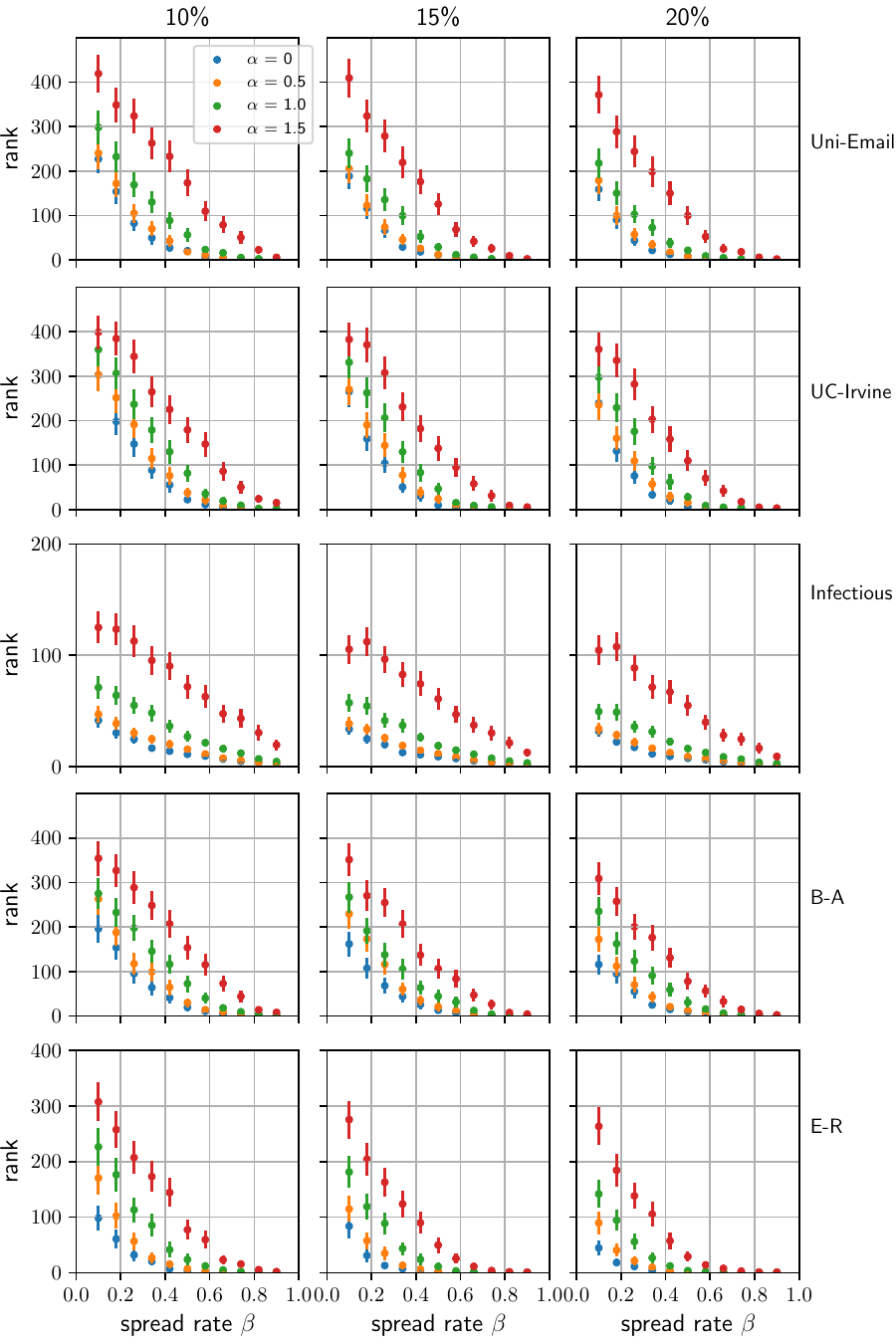}
            \caption{Average rank of true source decreases with the infection rate $\beta$ and increases with the IOL strength $\alpha$ (indicated in different colors).
            Rows correspond to different networks, and columns correspond to different observer density $\rho$.
            Each point is the average of 500 simulations.
            The error bars represent the extended standard error \((k=3)\).}
            \label{rank_beta_plot}
\end{figure}

\begin{figure}[H]
            \centering
            \includegraphics[width=0.85\textwidth]{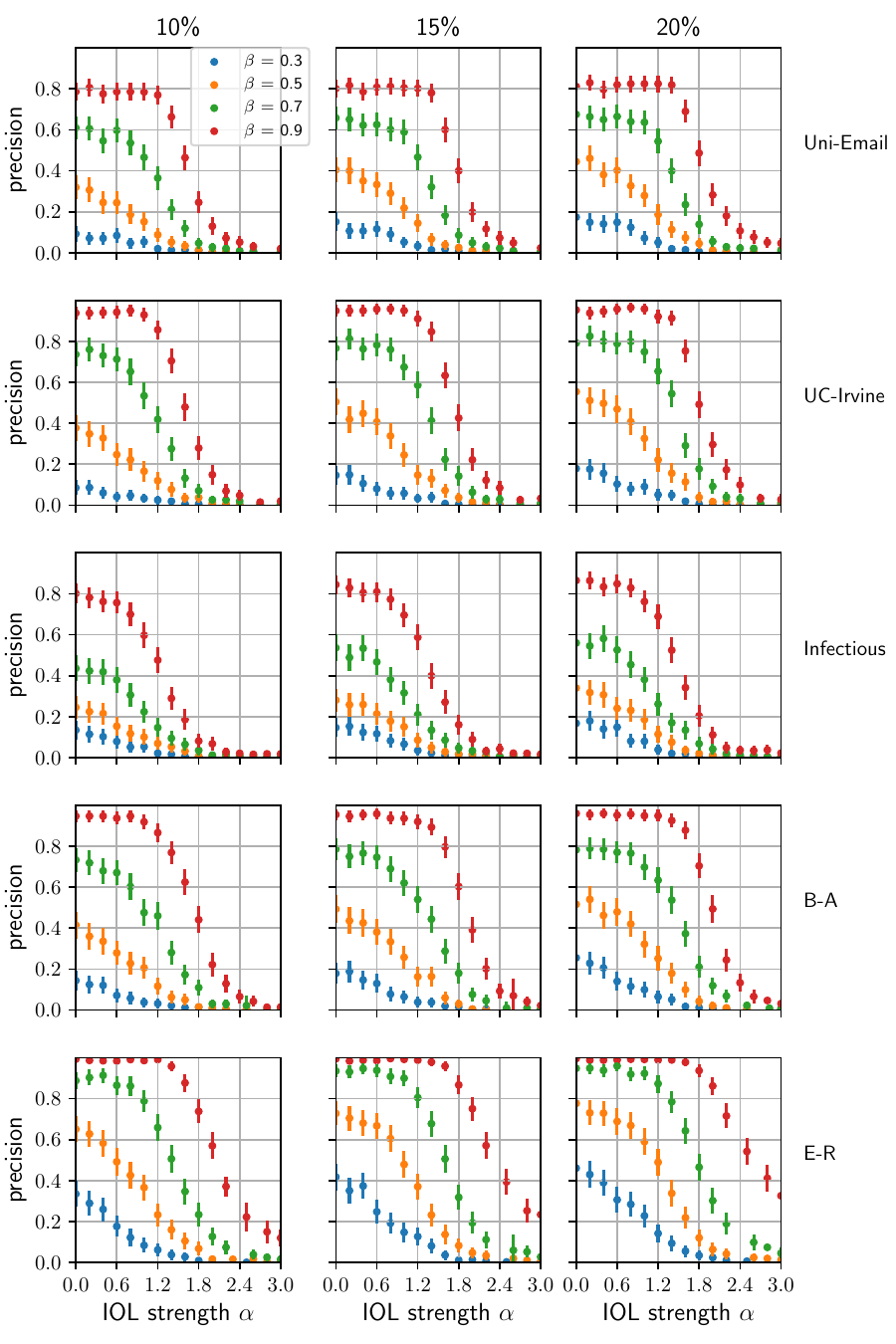}
            \caption{Average precision of source localization as a function of the IOL strength $\alpha$ for several levels of the infection rate $\beta$ (indicated in different colors).
            For $\beta \geq 0.7$, the average precision does not decrease monotonically, but it starts with a plateau.
            Rows correspond to different networks, and columns correspond to different observer density $\rho$.
            Each point is the average of 500 independent simulations.
            The error bars represent the extended standard error \((k=3)\).}
            \label{prec_alpha_plot}
\end{figure}

\begin{figure}[H]
            \centering
            \includegraphics[width=0.85\textwidth]{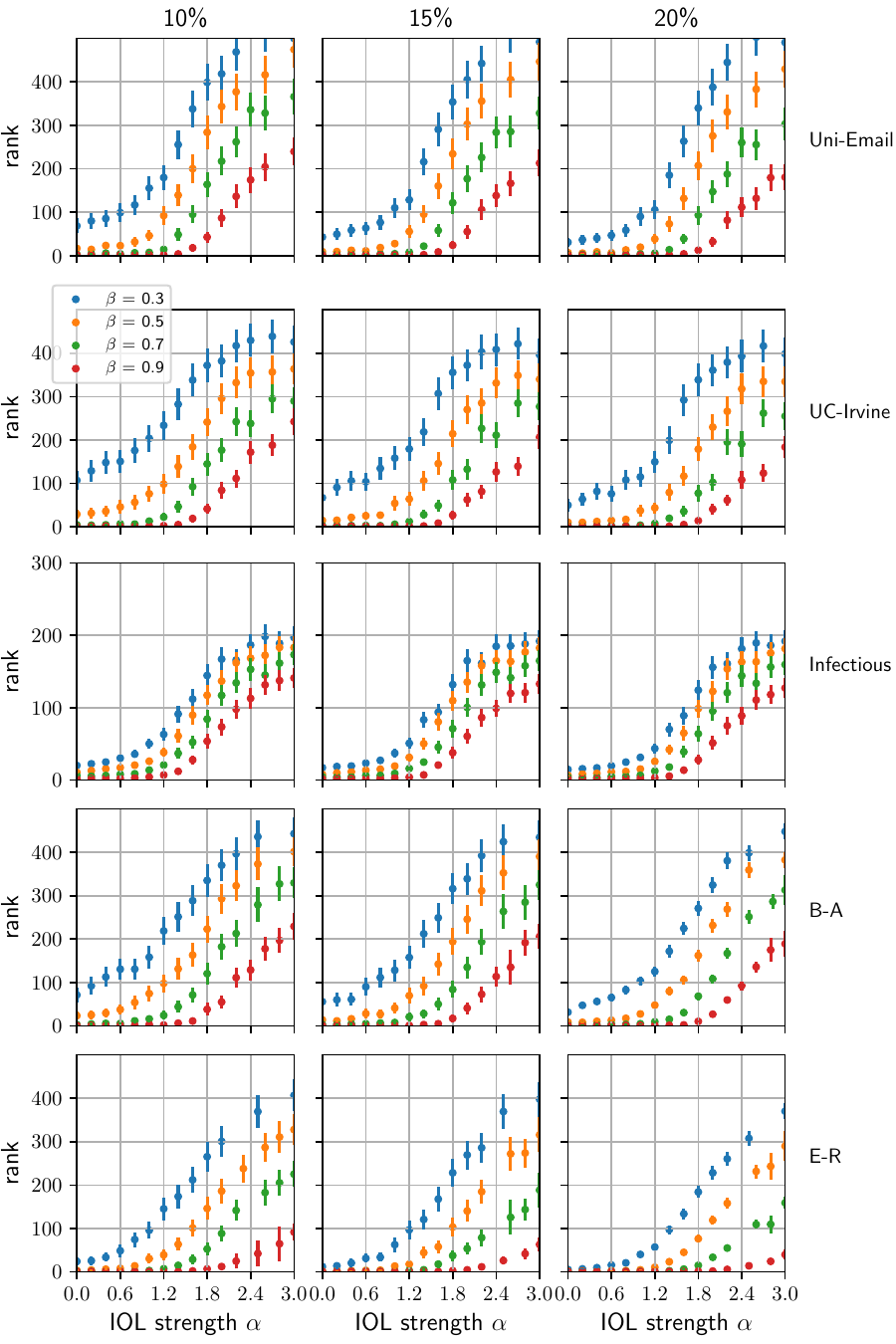}
            \caption{Average rank of true source as a function of the IOL strength $\alpha$ for several levels of the infection rate $\beta$ (indicated in different colors).
            For $\beta \geq 0.7$, the average rank does not increase monotonically, but it starts with a plateau.
            Rows correspond to different networks, and columns correspond to different observer density $\rho$.
            Each point is the average of 500 independent simulations.
            The error bars represent the extended standard error \((k=3)\).}
            \label{rank_alpha_plot}
\end{figure}

Key observations from Figs. \ref{prec_beta_plot}, \ref{rank_beta_plot} show that the source localization effectiveness increases as the spreading rate \(\beta\) increases.
Conversely, as seen in Figs. \ref{prec_alpha_plot}, \ref{rank_alpha_plot}, effectiveness decreases as the information overload strength $\alpha$ increases.
This decline in localization effectiveness is steeper at lower observer densities.
Moreover, the Email and UCIrvine networks are more resilient to IOL strength compared to the Infectious network, with UCIrvine having higher initial effectiveness.
This may be because these two networks have much lower clustering coefficients.
Comparing real networks to synthetic ones, it can be seen that the latter are more efficient, especially for E-R networks.

\begin{table}[H]
\centering
\caption{Topological parameters of the analyzed networks ($|V|$: number of vertices, $|E|$: number of edges, $\langle k \rangle$: average degree, $C$: clustering coefficient).}
\label{networks_table}
\begin{tabular}{lcccc}
\toprule
\textbf{Network} & $|V|$ & $|E|$ & $\langle k \rangle$ & $C$ \\
\midrule
Uni-Email & 1133 & 5451 & 9.62 & 0.046 \\
UCIrvine & 1020 & 6205 & 12.17 & 0.166 \\
Infectious & 410 & 2765 & 13.49 & 0.436 \\
\midrule
Erd\H{o}s-R\'{e}nyi & 1000 & 4000 & 8.00 & 0.008 \\
Barab\'{a}si-Albert & 1000 & 3997 & 7.99 & 0.026 \\
\bottomrule
\end{tabular}
\end{table}

\begin{figure}[H]
            \centering
            \includegraphics[width=1\textwidth]{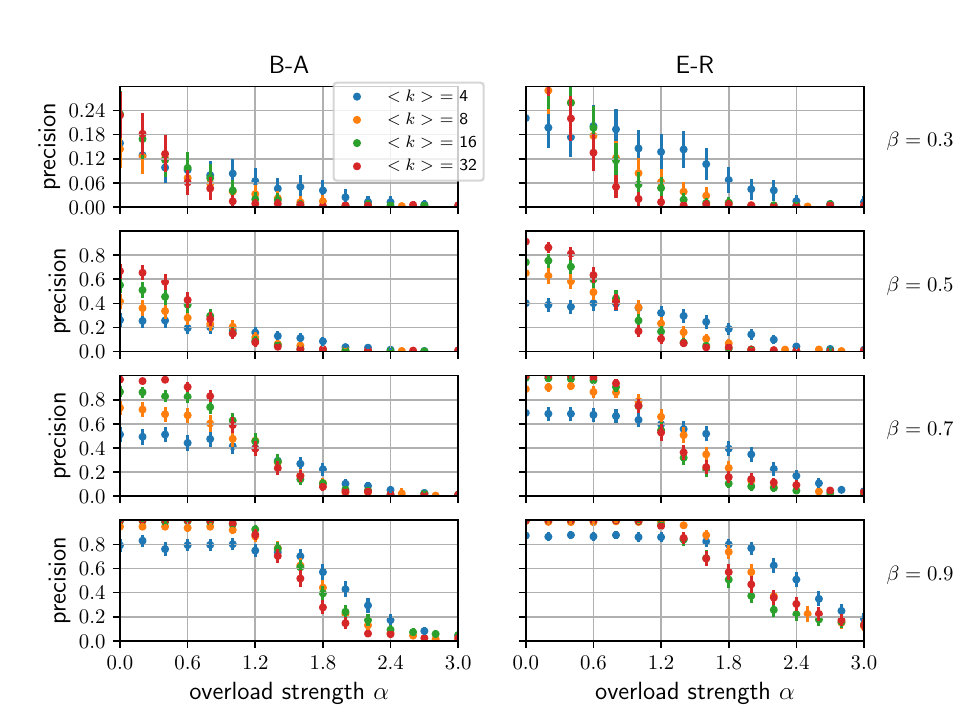}
            \caption{Average precision of source localization as a function of the IOL strength $\alpha$ for several values of the network average degree $\langle k \rangle$ (indicated in different colors).
            Networks with lower connection density are more resistant to information overload.
            Rows correspond to different infection rates $\beta$, and columns correspond to different network types. 
            Other parameters: number of nodes \(|V|=1000\), observer density \(\rho=10\%\).
            Each point is the average of 500 independent simulations.
            The error bars represent the extended standard error \((k=3)\).}
            \label{prec_k_avg_plot}
\end{figure}

\begin{figure}[H]
            \centering
            \includegraphics[width=1\textwidth]{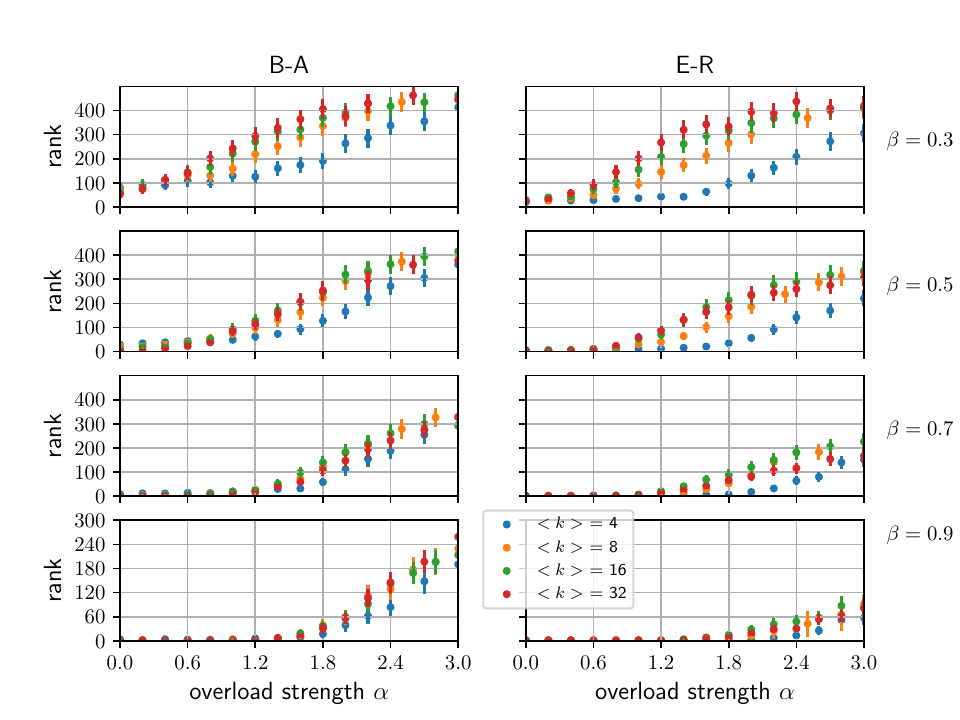}
            \caption{Average rank of true source as a function of the IOL strength $\alpha$ for several values of the network average degree $\langle k \rangle$ (indicated in different colors).
            Networks with lower connection density are more resistant to information overload.
            Rows correspond to different infection rates $\beta$, and columns correspond to different network types. 
            Other parameters: number of nodes \(|V|=1000\), observer density \(\rho=10\%\).
            Each point is the average of 500 independent simulations.
            The error bars represent the extended standard error \((k=3)\).}
            \label{rank_k_avg_plot}
\end{figure}

Figures \ref{prec_k_avg_plot} and \ref{rank_k_avg_plot} illustrate the effect of the network's density on the source localization quality. 
The simulations were conducted on the E-R and B-A networks, each with 1000 nodes and 100 observers.
The results indicate that the precision difference among different degree values is most pronounced for low and moderate spread rates.
In both types of networks for low values of IOL strength $\alpha$, the effectiveness is greater for higher average degrees $\langle k \rangle$.
However, as IOL strength increases, the less-dense networks surpass them in efficiency.
One can conclude that the impact of information overload on source localization effectiveness is greater for networks with a higher average degree $\langle k \rangle$.


\section{Discussion}
Source localization effectiveness clearly increases with the spreading rate \(\beta\) in all analyzed real-world and synthetic networks, as shown in Figs. \ref{prec_beta_plot} and \ref{rank_beta_plot}.
This positive correlation is consistent with previous findings for source localization within the standard SI model \citep{loc_alg_comp}.
However, our results provide a new insight: this improvement in localization quality persists across all analyzed values of the IOL strength \(\alpha\).
It suggests that the information overload phenomenon does not alter the fundamental relationship between the spreading rate and localization effectiveness. 

Conversely, Figs. \ref{prec_alpha_plot}-\ref{rank_alpha_plot} show that localization effectiveness decreases as the information overload strength \(\alpha\) increases.
It is indicated by both a decrease in average precision and an increase in average ranking.
It implies that stronger information overload makes it significantly harder to localize the true source using the Pearson correlation algorithm.
A conflict between the GFSIR propagation dynamics and the core assumption of the localization algorithm can explain this decline in effectiveness.
The algorithm assumes a linear relationship between a node's distance from the source and its infection time.
Information overload breaks this assumption.

\begin{figure}[H]
            \centering
            \includegraphics[width=0.9\textwidth]{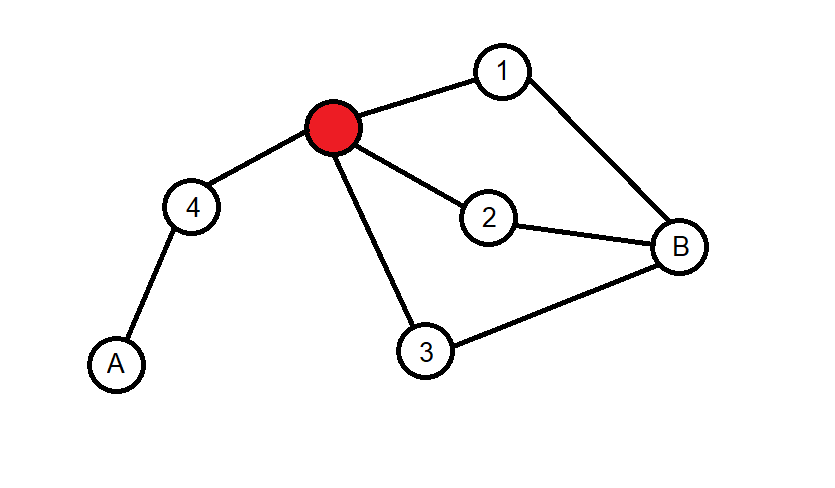}
            \caption{Surrounding of the source (red node) with observers labeled as $A$ and $B$}
            \label{src_surrounding}
\end{figure}

To illustrate this mechanism, consider the scenario in Fig. \ref{src_surrounding}. The source (red node) is surrounded by its neighbors, including two observers, \(A\) and \(B\).
For simplicity, we assume \(\beta=1\) (guaranteed spread) and \(\alpha>0\) (information overload is active).
In this case, at the first time step, nodes {1, 2, 3, 4} will be infected with certainty.
Crucially, although \(\alpha>0\), each of these nodes is infected by only one neighbor (the source).
Therefore, according to Equation \ref{fsir_state_eq} (where \(k_i(t)=1\)), information overload is not triggered at this step.
In the second time step, observer \(A\) will also be infected with certainty, as it receives information from only one node (node 4).
However, observer \(B\) faces a different situation: it is exposed to three infected nodes, {1, 2, 3}.
Due to information overload (where \(k_i(t)=3\)), the probability of infection from each node is reduced, resulting in a cumulative infection probability of less than one (see Equation \ref{fsir_state_eq}).
As a result, the expected activation time for observer \(B\) increases as the overload strength \(\alpha\) increases.
The Pearson algorithm scores the source by correlating the observer activation times \(t_o\) with their path distances from the source \(d(o,s)\).
This unexpected time delay for observer \(B\) (and other nodes exposed to multiple infected neighbors) breaks the linear distance-time relationship.
This mismatch weakens the correlation, reduces the source's score, and consequently harms the overall localization effectiveness.
This decline in effectiveness due to IOL is observed across all network types, spreading rates, and observer densities.
However, the effect is not always immediate; it is less visible at low \(\alpha\) values, especially when the spreading rate \(\beta\) is high.
This is particularly evident in the case of synthetic, Uni-Email, and UC-Irvine networks (Fig. \ref{prec_alpha_plot}), where precision remains high (maintains a plateau) for small \(\alpha\) values, only to decrease rapidly as \(\alpha\) crosses a certain threshold.
This resistance is less visible in the Infectious networks, suggesting that in networks with a high clustering coefficient, the negative effects of information overload may be stronger.

As expected, source localization quality is higher for the E-R networks than for the B-A networks.
Lower source localization effectiveness in scale-free networks has already been described for the SIR model \citep{b_a_loc_prec} and is evident from the initial precision values at $\alpha = 0$ in Fig. \ref{prec_alpha_plot}.
Regarding the impact of information overload, it can be concluded that E-R networks not only exceed B-A networks in initial effectiveness (when $\alpha=0$), but also have a higher $\alpha$ threshold at which effectiveness starts to drop.
The drop itself isn't as rapid, i.e., the source localization effectiveness is more resilient to IOL in E-R networks rather than in B-A networks.

To generalize from the studies presented above, we examined how IOL affects source localization in networks of varying density.
We can see in Figs. \ref{prec_alpha_plot}-\ref{rank_alpha_plot} that the network's susceptibility to IOL increases with increasing its average degree.
The reason for this is once again the localization algorithm's assumption about the linear spread of information.
As can be seen in Eq. \ref{fsir_state_eq}, the global value of the infection rate \(\beta\) is divided by \(k_i(t)^\alpha>1\), which means that the impact of \(\alpha\) on the infection probability depends on the number of infected neighbors.
Obviously, the number of infected neighbors is higher for denser networks.
Consequently, networks with a low mean degree are much more resistant to IOL.

\section{Conclusion and future work}
In this work, we modeled information spreading in selected real-world and synthetic networks, accounting for information overload using a fractional-order approach.
Our simulations demonstrate that the non-linear response to information overload significantly constrains the system's observability, negatively impacting the effectiveness of source localization across all analyzed topologies.

Our findings reveal a fundamental trade-off in the observability of complex networks under fractional constraints.
While higher network density typically facilitates information spread, our results demonstrate a non-linear `reversal' effect: in the presence of strong information overload, structural complexity acts as a source of topological noise that disrupts the linear distance-time correlation.
This transition suggests that the resilience of localization algorithms is not only a function of the network topology but is deeply coupled with the fractional dynamics of the agents' collective attention.

A significant strength of this study lies in the introduction of the GFSIR model, which provides a rigorous mathematical bridge between cognitive psychology -- specifically the information overload (IOL) phenomenon -- and the topological physics of spreading processes.
Unlike classical epidemiological models, our framework captures the fractional-order decay of transmission probability, offering a more nuanced representation of digital communication dynamics.
Furthermore, the identification of the density-dependent `reversal' effect provides a critical insight for network practitioners: it demonstrates that the traditional advantage of network density for information observability can become a liability under high fractional constraints.
This dual-layer analysis of both synthetic and real-world networks ensures that our findings are both theoretically grounded and practically relevant for misinformation tracing.

Despite these contributions, the present research is subject to several limitations that warrant acknowledgment.
First, our simulations assume a static, unweighted network topology, whereas real-world social platforms are inherently dynamic, with heterogeneous edges appearing and disappearing over time.
Second, the GFSIR model currently utilizes a homogeneous overload parameter \(\alpha\) for all nodes.
In reality, individuals exhibit varying cognitive capacities and susceptibility to information fatigue, suggesting that a distribution of \(\alpha\) values might yield more heterogeneous spreading patterns.
Finally, while Pearson’s correlation is a robust and efficient baseline for source localization, its reliance on linear distance-time assumptions means its performance is inherently bounded when non-linear fractional effects become dominant.

This study also opens several avenues for future research.
Firstly, further investigation into fractional-order propagation dynamics is required to better capture the response to information overload in heterogeneous populations.
One promising idea is to modify the GFSIR model to support a node-specific $\alpha$ parameter that captures individual differences in cognitive susceptibility.

An interesting concept would be to transfer the dynamics of the GFSIR model, which by definition describes multi-agent interactions, to higher-order networks.
Exploring such architectures is particularly pertinent, as higher-order interactions -- modeled via hypergraphs and simplicial complexes -- have recently emerged as a primary frontier in network science \citep{Dong2025, Guo2026}, offering a transformative paradigm for capturing the non-pairwise dependencies inherent in modern collective dynamics.

Another direction is to evaluate the robustness of additional localization algorithms against the information overload phenomenon.
Ultimately, extending the GFSIR framework to multi-source and multi-message scenarios would provide a more comprehensive model of the competitive dynamics of limited attention in real-world digital ecosystems.
In such a framework, distinct information items propagating simultaneously would compete for limited user attention, providing a more comprehensive model of information overload.

\section*{CRediT authorship contribution statement}
\textbf{Ignacy Czajkowski:} Software, Formal analysis, Investigation, Writing - Original Draft, \textbf{Robert Paluch:} Conceptualization, Methodology, Supervision, Writing - Review \& Editing.

\section*{Declaration of competing interest}
The authors declare that they have no known competing financial interests or personal relationships that could have appeared to influence the work reported in this paper.

\section*{Acknowledgments}
The research was funded by the Warsaw University of Technology within the Excellence Initiative: Research University (IDUB) programme.
Robert Paluch was also co-financed by the European Union under the Horizon Europe grant OMINO – Overcoming Multilevel INformation Overload (grant number 101086321, http://ominoproject.eu).
Views and opinions expressed are those of the authors alone and do not necessarily reflect those of the European Union or the European Research Executive Agency.
Neither the European Union nor the European Research Executive Agency can be held responsible for them.
Robert Paluch was also co-financed with funds from the Polish Ministry of Education and Science under the programme entitled International Co-Financed Projects.

\section* {Declaration of generative AI and AI-assisted technologies in the writing process}

While preparing this work, the authors used Grammarly and Gemini to improve language and readability. After using these tools, the authors reviewed and edited the content as needed. The authors take full responsibility for the publication's content.
\appendix
\section{Source code}
\label{app1}

The simulations were performed in Julia. The source code is available at \url{https://github.com/IgnacyCzajkowski/GFSIR_model/}.



\bibliographystyle{elsarticle-harv} 
\bibliography{refs}

@article{Holyst2024,
  title = {Protect our environment from information overload},
  volume = {8},
  ISSN = {2397-3374},
  DOI = {10.1038/s41562-024-01833-8},
  number = {3},
  journal = {Nature Human Behaviour},
  publisher = {Springer Science and Business Media LLC},
  author = {Hołyst,  Janusz A. and Mayr,  Philipp and Thelwall,  Michael and Frommholz,  Ingo and Havlin,  Shlomo and Sela,  Alon and Kenett,  Yoed N. and Helic,  Denis and Rehar,  Aljoša and Maček,  Sebastijan R. and Kazienko,  Przemysław and Kajdanowicz,  Tomasz and Biecek,  Przemysław and Szymanski,  Boleslaw K. and Sienkiewicz,  Julian},
  year = {2024},
  pages = {402–403}
}

@article{Guo2026,
title = {Robustness of interdependent directed hypergraphs with completely connected component},
journal = {{Chaos, Solitons \& Fractals}},
volume = {208},
pages = {118091},
year = {2026},
issn = {0960-0779},
doi = {https://doi.org/10.1016/j.chaos.2026.118091},
author = {Zhengao Guo and Yinzuo Zhou and Jie Zhou},
}

@article{Artime2024,
  title = {Robustness and resilience of complex networks},
  volume = {6},
  ISSN = {2522-5820},
  DOI = {10.1038/s42254-023-00676-y},
  number = {2},
  journal = {Nature Reviews Physics},
  publisher = {Springer Science and Business Media LLC},
  author = {Artime,  Oriol and Grassia,  Marco and De Domenico,  Manlio and Gleeson,  James P. and Makse,  Hernán A. and Mangioni,  Giuseppe and Perc,  Matjaž and Radicchi,  Filippo},
  year = {2024},
  month = jan,
  pages = {114–131}
}

@article{Ji2023,
title = {Signal propagation in complex networks},
journal = {Physics Reports},
volume = {1017},
pages = {1-96},
year = {2023},
issn = {0370-1573},
doi = {10.1016/j.physrep.2023.03.005},
author = {Peng Ji and Jiachen Ye and Yu Mu and Wei Lin and Yang Tian and Chittaranjan Hens and Matjaž Perc and Yang Tang and Jie Sun and Jürgen Kurths},
}

@article{Gosh2024,
title = {The eco-evolutionary dynamics of two strategic species: From the predator-prey to the innocent-spreader rumor model},
journal = {Journal of Theoretical Biology},
volume = {595},
pages = {111955},
year = {2024},
issn = {0022-5193},
doi = {10.1016/j.jtbi.2024.111955},
author = {Subrata Ghosh and Sourav Roy and Matjaž Perc and Dibakar Ghosh},
}

@article{Dong2025,
  title = {Adaptive rumor propagation and activity contagion in higher-order networks},
  volume = {8},
  ISSN = {2399-3650},
  DOI = {10.1038/s42005-025-02181-3},
  pages = {261},
  number = {1},
  journal = {Communications Physics},
  publisher = {Springer Science and Business Media LLC},
  author = {Dong,  Yafang and Huo,  Liang’an and Perc,  Matjaž and Boccaletti,  Stefano},
  year = {2025},
  month = jun 
}

@INPROCEEDINGS{Huang2025b,
  author={Huang, Tianyi and Yi, Jingyuan and Yu, Peiyang and Xu, Xiaochuan},
  booktitle={2025 8th International Conference on Advanced Algorithms and Control Engineering (ICAACE)}, 
  title={Unmasking Digital Falsehoods: A Comparative Analysis of LLM-Based Misinformation Detection Strategies}, 
  year={2025},
  volume={},
  number={},
  pages={2470-2476},
  doi={10.1109/ICAACE65325.2025.11020217}
}

@article{Lu2025,
author = {Lu, Yisong and Wei, Pengfei and Li, Jinsheng and Li, Qiong and Zhang, Xueqin},
doi = {10.1016/j.chaos.2025.116950},
issn = {0960-0779},
journal = {{Chaos, Solitons \& Fractals}},
pages = {116950},
title = {{Source identification in social networks based on graph transformer}},
volume = {200},
year = {2025}
}

@article{Fraszczak2025,
author = {Fr{\c{a}}szczak, Damian},
doi = {10.3390/e27090948},
journal = {Entropy},
number = {9},
pages = {948},
title = {{Identifying Network Propagation Sources Using Advanced Centrality Measures}},
volume = {27},
year = {2025}
}

@article{He2026,
author = {He,  Qiang and Yan,  Xin and Fang,  Hui and Chen,  Lu and Zhang,  Jie},
doi = {10.1109/TON.2025.3603678},
journal = {IEEE Transactions on Networking},
pages = {626--637},
publisher = {IEEE},
title = {{Tracing Epidemic Source of Dynamic Network Based on Graph Neural Network}},
volume = {34},
year = {2026}
}

@article{Ali2025,
author = {Ali, Syed Shafat and Rastogi, Ajay and Anwar, Tarique and Afzal, Syed and Rizvi, Murtaza and Yang, Jian and Wu, Jia and Member, Senior and Sheng, Quan Z},
doi = {10.1109/TKDE.2025.3567282},
journal = {IEEE Transactions on Knowledge and Data Engineering},
number = {8},
pages = {4620--4634},
publisher = {IEEE},
title = {{Generalized Local Prominence for Source Detection in Real-World Rumor Networks}},
volume = {37},
year = {2025}
}

@INPROCEEDINGS{Hu2025,
author={Hu, Wenyan and Ai, Jingxuan},
booktitle={2025 IEEE International Annual Conference on Complex Systems and Intelligent Science (CSIS-IAC)}, 
title={{Heterogeneous Network Rumor Propagation Considering Forgetting and Hot Topic Effects}}, 
year={2025},
volume={},
number={},
pages={1-6},
doi={10.1109/CSIS-IAC65538.2025.11162047}
}

@article{Huang2025,
author = {Huang, Da-wen and Wu, Wenjie and Bi, Jichao and Li, Junli and Gan, Chenquan and Zhou, Wei},
doi = {10.1016/j.ins.2024.121508},
issn = {0020-0255},
journal = {Information Sciences},
pages = {121508},
publisher = {Elsevier Inc.},
title = {{Timeliness-aware rumor sources identification in community-structured dynamic online social networks}},
volume = {689},
year = {2025}
}

@article{Lordan2020,
author = {Lordan, Oriol and Sallan, Jose M.},
doi = {10.1371/journal.pone.0242875},
issn = {19326203},
journal = {PLoS ONE},
number = {12},
title = {{Dynamic measures for transportation networks}},
volume = {15},
year = {2020}
}

@article{Ding2019,
author = {Ding, Rui and Ujang, Norsidah and Hamid, Hussain Bin and Manan, Mohd Shahrudin Abd and Li, Rong and Albadareen, Safwan Subhi Mousa and Nochian, Ashkan and Wu, Jianjun},
journal = {Networks and Spatial Economics},
doi = {10.1007/s11067-019-09466-5},
issn = {15729427},
number = {4},
pages = {1281--1317},
publisher = {Springer Science and Business Media LLC},
title = {{Application of Complex Networks Theory in Urban Traffic Network Researches}},
volume = {19},
year = {2019}
}

@article{b_a_loc_prec,
  title = {Fast and accurate detection of spread source in large complex networks},
  volume = {8},
  ISSN = {2045-2322},
  DOI = {10.1038/s41598-018-20546-3},
  number = {1},
  journal = {Scientific Reports},
  publisher = {Springer Science and Business Media LLC},
  author = {Paluch,  Robert and Lu,  Xiaoyan and Suchecki,  Krzysztof and Szymański,  Bolesław K. and Hołyst,  Janusz A.},
  year = {2018},
  month = feb 
}

@article{source_det_in_soc_net,
  title = {Source detection of rumor in social network – A review},
  volume = {9},
  ISSN = {2468-6964},
  DOI = {10.1016/j.osnem.2018.12.001},
  journal = {Online Social Networks and Media},
  publisher = {Elsevier BV},
  author = {Shelke,  Sushila and Attar,  Vahida},
  year = {2019},
  month = jan,
  pages = {30–42}
}

@article{eppler2004conceptual,
  title = {The Concept of Information Overload: A Review of Literature from Organization Science,  Accounting,  Marketing,  MIS,  and Related Disciplines},
  volume = {20},
  ISSN = {1087-6537},
  DOI = {10.1080/01972240490507974},
  number = {5},
  journal = {The Information Society},
  publisher = {Informa UK Limited},
  author = {Eppler,  Martin J. and Mengis,  Jeanne},
  year = {2004},
  month = nov,
  pages = {325–344}
}

@article{bawden2008information,
  title = {The dark side of information: overload,  anxiety and other paradoxes and pathologies},
  volume = {35},
  ISSN = {1741-6485},
  DOI = {10.1177/0165551508095781},
  number = {2},
  journal = {Journal of Information Science},
  publisher = {SAGE Publications},
  author = {Bawden,  David and Robinson,  Lyn},
  year = {2008},
  month = nov,
  pages = {180–191}
}

@article{inf_ovr_in_age,
  title = {Information overload in the information age: a review of the literature from business administration,  business psychology,  and related disciplines with a bibliometric approach and framework development},
  volume = {12},
  ISSN = {2198-2627},
  DOI = {10.1007/s40685-018-0069-z},
  number = {2},
  journal = {Business Research},
  publisher = {Springer Science and Business Media LLC},
  author = {Roetzel,  Peter Gordon},
  year = {2018},
  month = jul,
  pages = {479–522}
}

@article{fsir_model,
  title = {Competing for Attention in Social Media under Information Overload Conditions},
  volume = {10},
  ISSN = {1932-6203},
  DOI = {10.1371/journal.pone.0126090},
  number = {7},
  journal = {PLOS ONE},
  publisher = {Public Library of Science (PLoS)},
  author = {Feng,  Ling and Hu,  Yanqing and Li,  Baowen and Stanley,  H. Eugene and Havlin,  Shlomo and Braunstein,  Lidia A.},
  editor = {Chialvo,  Dante R.},
  year = {2015},
  month = jul,
  pages = {e0126090}
}

@article{corelation_algorythm,
  title = {Identifying the diffusion source in complex networks with limited observers},
  volume = {527},
  ISSN = {0378-4371},
  DOI = {10.1016/j.physa.2019.121267},
  journal = {Physica A: Statistical Mechanics and its Applications},
  publisher = {Elsevier BV},
  author = {Xu,  Shuaishuai and Teng,  Cong and Zhou,  Yinzuo and Peng,  Junhao and Zhang,  Yicheng and Zhang,  Zi-Ke},
  year = {2019},
  month = aug,
  pages = {121267}
}

@article{epidemic_spreading,
  title = {Epidemic spreading on adaptively weighted scale-free networks},
  volume = {74},
  ISSN = {1432-1416},
  DOI = {10.1007/s00285-016-1057-6},
  number = {5},
  journal = {Journal of Mathematical Biology},
  publisher = {Springer Science and Business Media LLC},
  author = {Sun,  Mengfeng and Zhang,  Haifeng and Kang,  Huiyan and Zhu,  Guanghu and Fu,  Xinchu},
  year = {2016},
  month = sep,
  pages = {1263–1298}
}

@article{zhu2016information,
  author={Zhu, Kai and Ying, Lei},
  journal={IEEE/ACM Transactions on Networking}, 
  title={{Information Source Detection in the SIR Model: A Sample-Path-Based Approach}}, 
  year={2016},
  volume={24},
  number={1},
  pages={408-421},
  doi={10.1109/TNET.2014.2364972}
}

@misc{information_overload_explain1,
  author = {Rogers, Paul and Puryear, Rudy and Root, James},
  title = {{Infobesity: The enemy of good decisions}},
  year = {2013},
  howpublished = {\url{https://www.bain.com/insights/infobesity- the-enemy-of-good-decisions/}},
  note = {Accessed: 2026-03-19}
}

@book{information_overload_explain2,
    author = {Wurman, R. S.},
    title = {Information Anxiety},
    publisher = {Doubleday},
    address = {New York},
    year = {1989}
}

@book{inf_ovr_book,
    author = {Gross, Bertram Myron},
    title = {The Managing Organizations: The Administrative Struggle},
    publisher = {The Free Press of Glencoe},
    address = "New York",
    year = {1964}
}

@book{inf_ovrload_hist,
  title = {Information Overload: An Introduction},
  ISBN = {9780190228637},
  DOI = {10.1093/acrefore/9780190228637.013.1360},
  journal = {Oxford Research Encyclopedia of Politics},
  publisher = {Oxford University Press},
  author = {Bawden,  David and Robinson,  Lyn},
  year = {2020},
  month = jun 
}

@article{sir_model_in_inf,
  title = {Generalization of Epidemic Theory: An Application to the Transmission of Ideas},
  volume = {204},
  ISSN = {1476-4687},
  DOI = {10.1038/204225a0},
  number = {4955},
  journal = {Nature},
  publisher = {Springer Science and Business Media LLC},
  author = {Goffman,  William and Newill,  Vaun A.},
  year = {1964},
  month = oct,
  pages = {225–228}
}

@article{source_det_alg_clas,
  title = {Identifying Propagation Sources in Networks: State-of-the-Art and Comparative Studies},
  volume = {19},
  ISSN = {1553-877X},
  DOI = {10.1109/comst.2016.2615098},
  number = {1},
  journal = {IEEE Communications Surveys \& Tutorials},
  publisher = {Institute of Electrical and Electronics Engineers (IEEE)},
  author = {Jiang,  Jiaojiao and Wen,  Sheng and Yu,  Shui and Xiang,  Yang and Zhou,  Wanlei},
  year = {2017},
  pages = {465–481}
}

@article{complex_networks_models,
  title = {Complex networks: Structure and dynamics},
  volume = {424},
  ISSN = {0370-1573},
  DOI = {10.1016/j.physrep.2005.10.009},
  number = {4–5},
  journal = {Physics Reports},
  publisher = {Elsevier BV},
  author = {Boccaletti,  Stefano and Latora,  Vito and Moreno,  Yamir and Chavez,  Mario and Hwang,  Dong-Uk},
  year = {2006},
  month = feb,
  pages = {175–308}
}

@article{networks_in_finance,
  title = {Lightning network: a second path towards centralisation of the Bitcoin economy*},
  volume = {22},
  ISSN = {1367-2630},
  DOI = {10.1088/1367-2630/aba062},
  number = {8},
  journal = {New Journal of Physics},
  publisher = {IOP Publishing},
  author = {Lin,  Jian-Hong and Primicerio,  Kevin and Squartini,  Tiziano and Decker,  Christian and Tessone,  Claudio J},
  year = {2020},
  month = aug,
  pages = {083022}
}

@article{networks_in_social,
  title = {A novel rumor detection with multi-objective loss functions in online social networks},
  volume = {213},
  ISSN = {0957-4174},
  DOI = {10.1016/j.eswa.2022.119239},
  journal = {Expert Systems with Applications},
  publisher = {Elsevier BV},
  author = {Wan,  Pengfei and Wang,  Xiaoming and Pang,  Guangyao and Wang,  Liang and Min,  Geyong},
  year = {2023},
  month = mar,
  pages = {119239}
}

@article{risk_dis_in_finance,
  title = {Measurement and contagion modelling of systemic risk in China’s financial sectors: Evidence for functional data analysis and complex network},
  volume = {90},
  ISSN = {1057-5219},
  DOI = {10.1016/j.irfa.2023.102913},
  journal = {International Review of Financial Analysis},
  publisher = {Elsevier BV},
  author = {Tian,  Sihua and Li,  Shaofang and Gu,  Qinen},
  year = {2023},
  month = nov,
  pages = {102913}
}

@article{delayed_development,
  title = {{Designing a sustainable-resilient-responsive supply chain network considering uncertainty in the COVID-19 era}},
  volume = {227},
  ISSN = {0957-4174},
  DOI = {10.1016/j.eswa.2023.120334},
  journal = {Expert Systems with Applications},
  publisher = {Elsevier BV},
  author = {Moadab,  Amirhossein and Kordi,  Ghazale and Paydar,  Mohammad Mahdi and Divsalar,  Ali and Hajiaghaei-Keshteli,  Mostafa},
  year = {2023},
  month = oct,
  pages = {120334}
}

@article{rumors_and_misinf_spread,
  title = {The spread of true and false news online},
  volume = {359},
  ISSN = {1095-9203},
  DOI = {10.1126/science.aap9559},
  number = {6380},
  journal = {Science},
  publisher = {American Association for the Advancement of Science (AAAS)},
  author = {Vosoughi,  Soroush and Roy,  Deb and Aral,  Sinan},
  year = {2018},
  month = mar,
  pages = {1146–1151}
}

@article{si_model,
    author = {Anderson, R. M. and May, R. M.},
    title = {Infectious diseases of humans: dynamics and control},
    journal = {Oxford University Press},
    doi = {10.1093/oso/9780198545996.001.0001},
    year = {1991}
}

@article{sis_model,
  title = {A contribution to the mathematical theory of epidemics},
  volume = {115},
  ISSN = {2053-9150},
  DOI = {10.1098/rspa.1927.0118},
  number = {772},
  journal = {Proceedings of the Royal Society of London. Series A,  Containing Papers of a Mathematical and Physical Character},
  publisher = {The Royal Society},
  author = {Kermack,  William Ogilvy and McKendrick,  A. G.},
  year = {1927},
  month = aug,
  pages = {700–721}
}

@article{multi_source_det,
  title = {Locating Multi-Sources in Social Networks With a Low Infection Rate},
  volume = {9},
  ISSN = {2334-329X},
  DOI = {10.1109/tnse.2022.3153968},
  number = {3},
  journal = {IEEE Transactions on Network Science and Engineering},
  publisher = {Institute of Electrical and Electronics Engineers (IEEE)},
  author = {Zhu,  Peican and Cheng,  Le and Gao,  Chao and Wang,  Zhen and Li,  Xuelong},
  year = {2022},
  month = may,
  pages = {1853–1865}
}

@article{multi_source_det2,
    author = {Fu, L. and Shen, Z. S. and Wang, W. X. and Fan, Y. and Di, Z. R.},
    title = {Multi-source localization on complex networks with limited observers.},
    journal = {EPL},
    volume = {113},
    pages = {18006},
    doi = {10.1209/0295-5075/113/18006},
    year = {2016}
}

@article{ptv_algorithm,
  title = {Locating the Source of Diffusion in Large-Scale Networks},
  volume = {109},
  ISSN = {1079-7114},
  DOI = {10.1103/physrevlett.109.068702},
  number = {6},
  journal = {Physical Review Letters},
  publisher = {American Physical Society (APS)},
  author = {Pinto,  Pedro C. and Thiran,  Patrick and Vetterli,  Martin},
  year = {2012},
  month = aug 
}

@article{s_z_algorithm,
  title = {Rumors in a Network: Who’s the Culprit?},
  volume = {57},
  ISSN = {1557-9654},
  DOI = {10.1109/tit.2011.2158885},
  number = {8},
  journal = {IEEE Transactions on Information Theory},
  publisher = {Institute of Electrical and Electronics Engineers (IEEE)},
  author = {Shah,  Devavrat and Zaman,  Tauhid},
  year = {2011},
  month = aug,
  pages = {5163–5181}
}

@article{sso_framework,
title = {Social media overload, exhaustion, and use discontinuance: Examining the effects of information overload, system feature overload, and social overload},
journal = {Information Processing and Management},
volume = {57},
number = {6},
pages = {102307},
year = {2020},
issn = {0306-4573},
doi = {10.1016/j.ipm.2020.102307},
author = {Fu, S. and Li, H. and Liu, Y. and Pirkkalainen, H. and Salo, M.},
}

@article{soc_econ_stat_inf_ovr,
title = {Contextualized impacts of an infodemic on vaccine hesitancy: The moderating role of socioeconomic and cultural factors},
journal = {Information Processing and Management},
volume = {59},
number = {5},
pages = {103013},
year = {2022},
issn = {0306-4573},
doi = {10.1016/j.ipm.2022.103013},
author = {Lin, F. and Chen, X. and Cheng, E. W.},
}

@article{ml_for_inf_relevance_class,
title = {Rapid relevance classification of social media posts in disasters and emergencies: A system and evaluation featuring active, incremental and online learning},
journal = {Information Processing and Management},
volume = {57},
number = {1},
pages = {102132},
year = {2020},
issn = {0306-4573},
doi = {10.1016/j.ipm.2019.102132},
author = {Kaufhold, M.-A. and Bayer, M. and Reuter, C.},
}

@inproceedings{real_networks,
author = {Kunegis, J\'{e}r\^{o}me},
title = {{KONECT: the Koblenz network collection}},
year = {2013},
isbn = {9781450320382},
publisher = {Association for Computing Machinery},
address = {New York, NY, USA},
doi = {10.1145/2487788.2488173},
booktitle = {Proceedings of the 22nd International Conference on World Wide Web},
pages = {1343–1350},
numpages = {8},
location = {Rio de Janeiro, Brazil},
series = {WWW '13 Companion}
}

@article{email_network1,
title = {Self-similar community structure in a network of human interactions.},
journal = {Phys. Rev. E Stat. Nonlinear Soft Matter Phys.},
volume = {68},
number = {065103},
year = {2003},
doi = {10.1103/PhysRevE.68.065103},
author = {Guimerà, R. and Danon, L. and Díaz-Guilera, A. and Giralt, F. and Arenas, A.},
}

@article{ucirving_network,
  title = {Clustering in weighted networks},
  volume = {31},
  ISSN = {0378-8733},
  DOI = {10.1016/j.socnet.2009.02.002},
  number = {2},
  journal = {Social Networks},
  publisher = {Elsevier BV},
  author = {Opsahl,  Tore and Panzarasa,  Pietro},
  year = {2009},
  month = may,
  pages = {155–163}
}

@article{infectious_network,
  title = {{What’s in a crowd? Analysis of face-to-face behavioral networks}},
  volume = {271},
  ISSN = {0022-5193},
  DOI = {10.1016/j.jtbi.2010.11.033},
  number = {1},
  journal = {Journal of Theoretical Biology},
  publisher = {Elsevier BV},
  author = {Isella,  Lorenzo and Stehlé,  Juliette and Barrat,  Alain and Cattuto,  Ciro and Pinton,  Jean-Fran\c{c}ois and Van den Broeck,  Wouter},
  year = {2011},
  month = feb,
  pages = {166–180}
}

@article{loc_alg_comp,
  title = {Comparison of observer based methods for source localisation in complex networks},
  volume = {12},
  ISSN = {2045-2322},
  DOI = {10.1038/s41598-022-09031-0},
  number = {1},
  journal = {Scientific Reports},
  publisher = {Springer Science and Business Media LLC},
  author = {Gajewski,  L. G. and Paluch,  Robert and Suchecki,  Krzysztof and Sulik,  Adam and Szymanski,  Boleslaw K. and Hołyst,  Janusz A.},
  year = {2022},
  month = mar 
}





\end{document}